\documentclass[aps,preprint,prl,showpacs,superscriptaddress]{revtex4}
\usepackage{graphicx}
\usepackage{amsmath,amssymb,latexsym,color,amsfonts}
\bibstyle{apsrev.bib}

\newcommand{\beq}{\begin{equation}}
\newcommand{\eeq}{\end{equation}}
\begin{document}


\title{THz-pump -- THz-probe spectroscopy of semiconductors at high field strengths}

\author{Matthias C. Hoffmann}
\email[mch@mit.edu]{}
\affiliation{Massachusetts Institute of
Technology}
\author{J\'{a}nos Hebling}
\affiliation{Department of Physics,University of P\'{e}cs,
Hungary}
\author{Harold Y. Hwang}
\author{Ka-Lo Yeh}
\author{Keith A. Nelson}
\affiliation{Massachusetts Institute of Technology}

\date{\today}

\begin{abstract}
Pumping n-type  GaAs and InSb with ultrafast THz pulses having
intensities higher than 150 MW/cm$^2$ shows strong  free carrier
absorption saturation at temperatures of 300 K and 200 K
respectively. If the energy imparted to  the carriers exceeds the
bandgap, impact ionization processes can occur. The dynamics of
carrier cooling in GaAs and impact ionization in InSb were
monitored using THz-pump/THz probe spectroscopy which provides
both sub-bandgap excitation and probing, eliminating any direct
optical electron-hole generation that complicates the evaluation
of results in optical pump/THz probe experiments.

\end{abstract}


\maketitle

\section{Introduction}

The study of hot carrier effects plays a central role in the
advancement of semiconductor science. Properties of hot carriers
are influenced both by the interaction between carriers and that
between the lattice and the carriers. Information about these
scattering processes, which determine high-field transport
phenomena in semiconductors, is highly valuable because they
constitute the basis of many ultrafast electronic and
optoelectronic devices. With the advent of ultrafast lasers, the
dynamics of hot carrier effects have been studied intensively on
the picosecond and femtosecond timescales. In a typical
experiment, new carriers are generated by an optical pump beam
with a photon energy, $\epsilon$$_{ph}$, above the bandgap,
$\epsilon$$_{g}$, of the semiconductor. The newly generated
electrons and holes share the excess energy
$\epsilon$$_{ph}$-$\epsilon$$_{g}$ in a ratio inversely
proportional to their effective masses and subsequently undergo
cooling processes mediated by carrier-carrier and carrier phonon
scattering. The dynamics of these cooling processes can then be
monitored by short probe pulses of appropriate wavelength. This
technique has successfully been applied to a large range of
semiconductors and semiconductor nanostructures and a substantial
amount of basic information was collected in this way
\cite{shah1999}.

With photo excitation, an equal number of electrons and holes are
created. Hence, most ultrafast optical studies are performed in
the presence of a plasma containing both types of carriers.
Experimental observations can be complicated by the dynamics that
involve both electrons and holes. When properties like the
coupling to the phonons and carrier-carrier scattering are
different for the two components and the density of carriers is
time dependent, a direct interpretation of the experimental data
is further hindered. Near-IR pump/THz probe experiments also
suffer from this shortcoming
\cite{vanexter1990,ralph1996,nuss1987}.

In this article we report results from THz-pump/THz-probe
measurements used to monitor the inter- and intra-valley dynamics
of extremely hot free electrons in GaAs and the generation of new
carriers by impact ionization in InSb. THz electric field
strengths up to 150 kV/cm at the surface of the semiconductor were
achieved.  Fields of this magnitude and correspondingly hot
electrons are present inside different fast semiconductor devices
like Gunn-diodes and avalanche photo diodes. Electron heating by
the THz pulse and strong intervalley scattering cause a large
fraction of the electrons to scatter out from the initial lowest
energy conduction band valley into side valleys.  Based on a rough
estimate, electrons can reach an average energy on the order of 1
eV. Since free-carrier absorption is proportional to carrier
concentration and carrier mobility, and because different
conduction band valleys usually have significantly different
mobilities from each other, the change in the distribution of
electrons amongst the different valleys causes a change in THz
absorption. This behaviour has been observed with nanosecond
transmission experiments with field strengths up to 10 kV/cm
\cite{mayer1986} and was monitored via the absorption of a delayed
probe THz pulse.

\section{Experimental technique}

\begin{figure}
  \includegraphics[width=10cm]{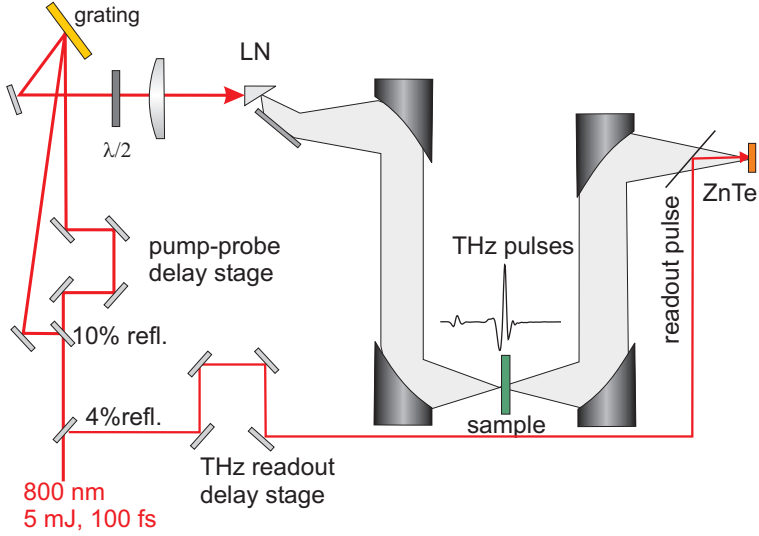}\\
  \caption{(color online) Schematic illustration of the experimental setup. Collinear THz pulses are generated by tilted pulse front excitation in LiNbO3  (LN) and detected electro-optically. See text for details.}\label{figure1}
\end{figure}

The experimental setup shown in Figure \ref{figure1} was used to
elucidate the dynamics of hot carriers in doped semiconductors. We
generated single-cycle THz pulses by optical rectification of
pulses from a femtosecond laser using the tilted-pulse-front
method \cite{hebling2002}. This technique uses a tilted intensity
front of a femtosecond laser pulse to achieve velocity matching of
the phonon polaritons inside lithium niobate and the femtosecond
laser light \cite{feurer2007}. The tilted pulse front enables the
use of high optical pulse energy to build up large THz field
amplitudes while averting unwanted nonlinear optical effects. THz
pulse energies greater than 10 $\mu$J can be achieved using this
method \cite{yeh2007}. Because the angle of the pulse front tilt
can be arbitrarily chosen, the method can also be adapted for
other materials and wavelengths \cite{hoffmann2007}. For our
experiments a regeneratively amplified titanium sapphire laser
system with 6 mJ pulse energy and 100 fs pulse duration at a
repetition rate of 1 kHz was used. The optical beam was split
using a 10:90 beamsplitter into two parts that were recombined
under a small angle at the same spot on the grating. The 10\% part
was passed through a chopper wheel (not shown in figure) and was
used to generate the THz probe. The 90\% part was variably
delayed, and used to generate the THz pump pulse.
The single-cycle THz pulses were focused onto the sample using a
90-degree off-axis parabolic mirror pair with 190 and 75 mm focal
lengths. The ratio of focal lengths allows us to reduce the beam
diameter at the sample to 1 mm. Another off-axis parabolic mirror
pair with focal lengths of 100 and 190 mm was used to image the
sample plane onto the ZnTe detector crystal for electro-optic
sampling of the THz field using balanced detection and a lock-in
amplifier \cite{zhang1995}. Because larger than 2 $\pi$ phase
shifts are routinely observed with thicker ZnTe crystals, we used
a ZnTe compound detector with an active layer of 0.1 mm and a
total thickness of 1.1 mm  to ensure the linearity of the detected
signal and to eliminate THz pulse reflections  within the crystal
\cite{turchinovich2007}. Selective chopping of the probe beam
provided excellent suppression of the pump pulse. Spectral
analysis of our THz pump-probe results was conducted in the 0.2 to
1.6 THz range where the spectral amplitude was sufficiently high.
A pair of wiregrid polarizers was used to attenuate the THz pulses
for intensity-dependent studies. We measured the THz fields
\textit{E}(\textit{t}) that reached the ZnTe crystal with and
without the sample in the beam path and calculated the effective
absorption coefficient

\begin{equation}
\alpha_{\mathrm{eff}}=-\frac{1}{d}\ln\left(T^2\cdot\frac{\int_0^{t_{\mathrm{max}}}
E^2_{\mathrm{sam}}(t)dt}{\int_0^{t_{\mathrm{max}}}E^2_{\mathrm{ref}}(t)dt}\right)
\end{equation}

where \textit{d} is the sample thickness, \textit{t}$_{max}$ is
the time window of the measurement and \textit{T} is a factor
accounting for reflection losses at the sample surfaces. The
quantity $\alpha$$_{eff}$ is equivalent to the energy absorption
coefficient averaged over our bandwidth. Referencing the recorded
data against the electric field measured without the sample
enables us to compensate nonlinear effects within the LN
generation crystal to a large degree. Only in a time interval of
$\pm$1 ps around the overlap time of the pump and probe does the
data quality become unreliable.

\textbf{}

The same setup can be reconfigured for intensity dependent
transmission studies by simply blocking the optical pulse used to
generate the THz probe beam, chopping the portion used to create
the THz pump, and changing the angle between the first and second
THz wiregrid polarizers to vary the intensity of the THz pump
pulse.

\section{Hot electron dynamics in n-type GaAs}

\subsection{Saturation of free-carrier absorption}

Free-carrier absorption in the THz range, which can be described
by the Drude model, is readily observed in bulk doped
semiconductors \cite{vanexter1990,huggard2000}.  Linear THz-TDS
measurements of n-type GaAs revealed the Drude behavior of GaAs
that is dependent on both the doping concentration and mobility of
the sample.
\begin{figure}
  \includegraphics[width=15cm]{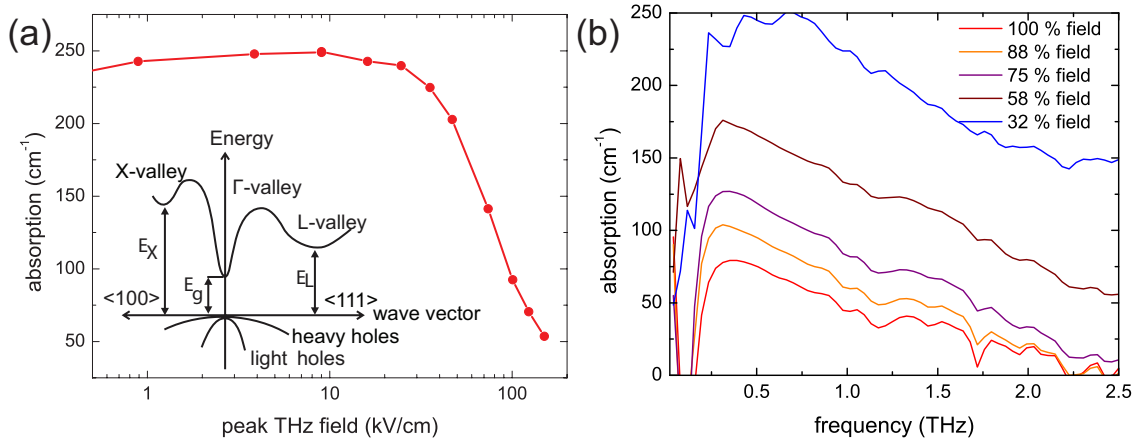}\\
  \caption{(color online) (a) Average THz absorption in n-type GaAs for peak fields between 1 and 150 kV/cm, the inset shows the simplified band structure of GaAs. (b) Frequency resolved absorption spectra at selected field strengths.}\label{figure2}
\end{figure}

In order to investigate free-carrier absorption in strong THz
fields, nonlinear THz transmission measurements were performed on
a 450 $\mu$m thick, n-type GaAs wafer with a carrier concentration
of 8$\times$10$^{15}$ cm$^{-3}$ at 300 K. THz pulse energies and
peak fields of up to 2 $\mu$J and 150 kV/cm respectively were
used. Figure \ref{figure2}a shows the observed averaged THz
absorption between 0.35 and 1.5 THz for a range of peak fields. At
THz fields smaller than 10 kV/cm, the absorption approaches the
values obtained in linear THz transmission experiments conducted
using a spectrometer with photoconductive switches.

Starting at peak fields of 30 kV/cm, we observe a drastic drop in
the absorption coefficient. At fields larger than 100 kV/cm, the
saturation effect appears to level off slightly.

Figure \ref{figure2}b shows the corresponding absorption spectra
for various selected field strengths above 50 kV/cm. Even though
the overall spectrum does not change in shape, the absorption
drops uniformly over a broad frequency range.

The decrease in absorption can be explained qualitatively by the
change of carrier mobility due to the acceleration of the free
electrons in the conduction band by the electric field of the THz
pulse.  Taking into account energy relaxation during the 1 ps pump
pulse duration, we estimate the average carrier energy just after
the pump pulse to be in the 0.5--1.3 eV range, exceeding the
energy necessary for carriers to cross into the side valleys.
These values are consistent with Monte-Carlo simulations that show
that electrons can reach ballistic velocities of up to 10$^{8
}$cm/s within 30 fs in external electric fields
\cite{shichijo1981}. The ballistic acceleration competes with
phonon scattering processes after 30 fs, leading to an average
heating of the electron gas on a timescale faster than the THz
pulse duration of 1 ps.

Electrons with these high energies can scatter into satellite
valleys (L and X) of the conduction band which have an energy
separation from the bottom of the zone center ($\Gamma$-valley) of
0.29 eV and 0.48 eV respectively. The effective mass at the zone
center is 0.063 m$_{0}$ leading to high mobility and high THz
absorption. The density of states effective mass in the side
valleys is much higher (0.85 m$_{0}$) resulting in much smaller
mobility.

In addition to the difference in mobilities among the initial and
side valleys, the nonparabolicity of the valleys can also result
in a decrease in mobility and a concomitant decrease in THz
absorption that accompanies the increase in carrier kinetic energy
within a single valley. Energy-dependent effective mass and
nonparabolicity parameters reported for GaAs \cite{shichijo1981}
indicate that adding 0.3 eV of kinetic energy to electrons in the
lowest-energy valleys reduces their THz absorption by 2/3.

\subsection{Time resolved absorption measurements}

With simple intensity dependent transmission measurements we are
unable to observe relaxation of the excited carriers back into the
$\Gamma$ valley after the strong heating of the electron gas by
the THz pulse. This cooling process can only be observed via time
resolved measurements. In particular, different relaxation times
are expected for inter- and intra-valley relaxations,
necessitating the use of the THz-pump/THz-probe technique
described in the earlier experimental section.
\begin{figure}
  \includegraphics[width=15cm]{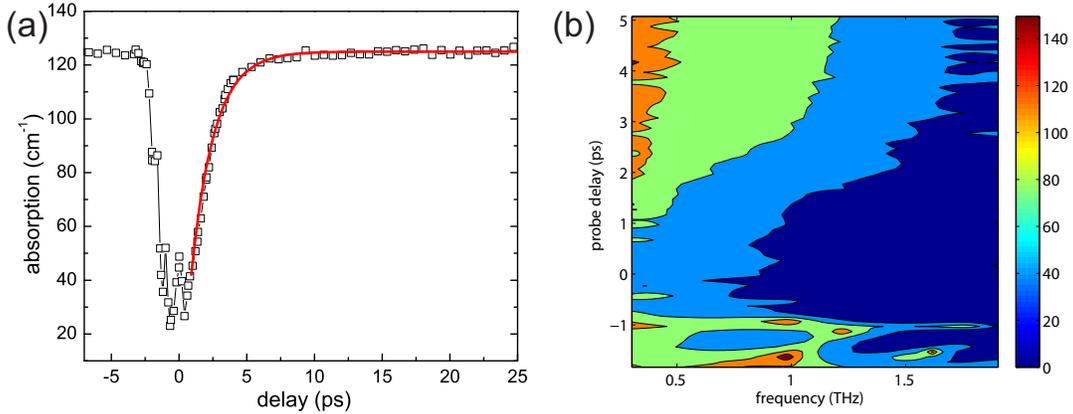}\\
  \caption{(color online): (a) The recovery of the absorption as a function of THz probe delay after the arrival of a strong THz pump pulse at t=0. (b) Frequency dependent absorption coefficient (cm-1)  as a function of probe delay times, revealing the relaxation of the excited carriers back to their equilibrium Drude-like behavior 5 ps after the arrival of the pump pulse.}\label{figure3}
\end{figure}

Figure \ref{figure3}a shows the time resolved absorption upon THz
excitation of the n-type GaAs sample. The equilibrium
frequency-averaged absorption of 125 cm$^{-1}$ drops below 30
cm$^{-1}$ immediately after the arrival of the THz pump pulse. The
drop in absorption occurs on the same timescale as the time
resolution provided by our pulses. A complete recovery of the
absorption is reached after 7 ps.  An exponential fit of the
absorption (solid line in Figure \ref{figure3}a) yields a carrier
relaxtion time of $\tau$$_{r}$ = 1.9 ps. In the frequency domain,
(shown in Figure \ref{figure3}b), we observe a slow recovery of
the Drude absorption until the initial absorption is restored. The
observation that the  relaxation time from the L-valley is much
larger than the scattering time into it can be explained by the
fact that it is governed by the relaxation rate from the upper
$\Gamma$ state to the lower $\Gamma$ state, and only those
carriers in the lower $\Gamma$ state are allowed to contribute to
the signal. Rate equations for the intervalley scattering have
been employed by Stanton \cite{stanton1992}.

\begin{figure}
  \includegraphics[width=8cm]{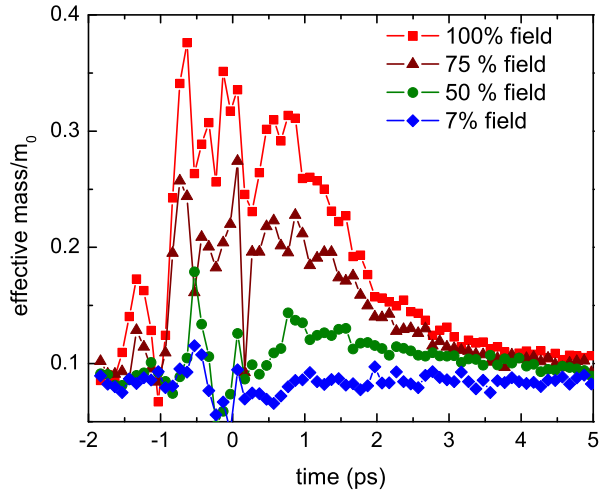}\\
  \caption{(color online): Average effective mass, relative to the electron mass, obtained from Drude-fits to the absorption spectra for different pump field strengths.}\label{figure4}
\end{figure}

Similar to the field dependent nonlinear transmission measurements
shown in Figure \ref{figure2}a, the time resolved absorption
spectra in our pump-probe measurements can be fit to a simple
Drude-model. Keeping the carrier concentration at the constant
value obtained from Hall measurements by the manufacturer, we can
extract an averaged effective mass. The result of the time
evolution of this quantity for different pump strengths is shown
in Figure \ref{figure4}. At high field strengths the average
effective mass exceeds a value of 0.3 m$_{0}$$_{ }$suggesting that
a sizeable fraction of the carriers is scattered into the side
valleys with higher mass. Qualitatively we also observe a rise in
the effective scattering time, although the data quality does not
allow us to make quantitative conclusions.

\section{THz-induced impact ionization in InSb}

Indium antimonide (InSb) has the highest electron mobility and
saturation velocity of all known semiconductors, enabling the
fabrication of transistors with extremely high switching speeds
\cite{ashley2004}. The elucidation of carrier dynamics in InSb on
the ultrashort timescale is hence of great fundamental and
technological relevance.  The material is a direct semiconductor
with a bandgap of 170 meV at room temperature \cite{littler1985},
making it well-suited for applications in infrared sensors
covering the wavelength range from 1  $\mu$m to 5 $\mu$m
\cite{avery}.

Impact ionization by high electric fields is a well known
phenomenon in InSb \cite{ancker1972}. The effect is determined by
the probability that an electron will gain enough energy from the
driving field to cross the ionization threshold. This is usually
observed at relatively low DC fields of several hundred V/cm where
avalanche effects play an important role. Strong THz fields can
directly achieve impact ionization on the picosecond time scale
\cite{wen2008} while avoiding additional experimental
complications by the use of photons with energies well below the
bandgap.

Because of the small bandgap, undoped InSb has a high intrinsic
carrier concentration of 2$\times$10$^{16}$ cm$^{-3}$at room
temperature. Even at 200 K the absorption still exceeds 200
cm$^{-1}$ at frequencies below 1 THz. At temperatures of 80K, the
intrinsic carrier concentration is on the order of 10$^{9}$$^{
}$cm$^{-3 }$and the remaining carrier concentration is dominated
by impurities.

In intensity dependent transmission measurements at high intrinsic
carrier concentrations (200 K), we observe a behavior similar to
that of GaAs (Figure \ref{figure2}a) with a drop in average
absorption to 50\% of the low intensity value. At low carrier
concentration of 2$\times$10$^{14}$ cm$^{-3}$ (80 K) we observe a
small rise in THz absorption at high fields in transmission
measurements, similar to the results reported in \cite{wen2008},
hinting at new carrier generation by the impact ionization
process.

\begin{figure}
  \includegraphics[width=8cm]{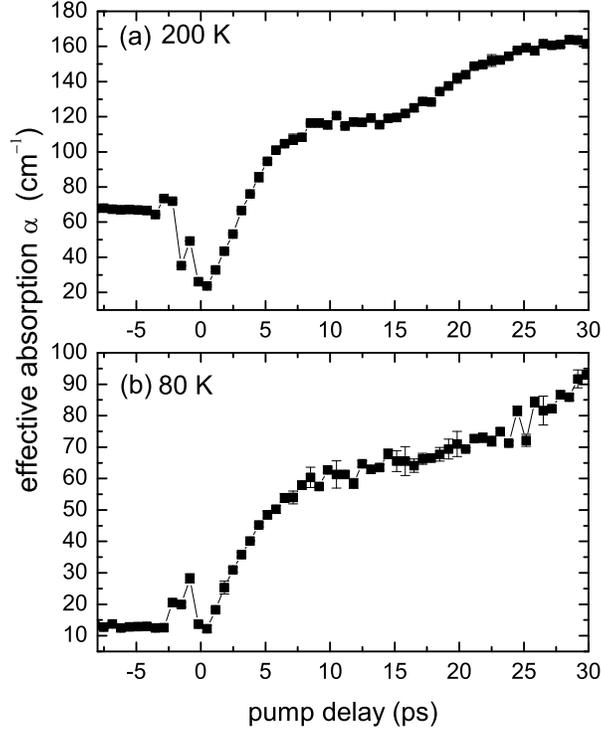}\\
  \caption{Time-resolved THz absorption of doped InSb at 200K and 80K after excitation by a
2 $\mu$J THz pulse. }\label{figure5}
\end{figure}

In order to understand the dynamics of carrier heating and
subsequent impact ionization we carried out pump-probe
measurements by the method described above. Figure \ref{figure5}
shows time resolved absorption data for n-type Te-doped InSb with
a carrier concentration of 2.0$\times$10$^{15}$ cm$^{-3}$$^{ }$at
77 K. The mobility as specified by the manufacturer was
2.5$\times$10$^{5}$ cm$^{2}$/Vs. The THz fields were polarized
parallel to the (100) axes of the crystals. At sample temperatures
of 200 K and 80 K the absorption increases by roughly the same
amount (80-90 cm$^{-1}$) after 30 ps.   At 200 K, there is an
initial dip in absorption before a subsequent rise after a few ps,
confirming the findings in our transmission measurements. No such
initial decrease is observed at 80 K. The difference in the early
time absorption behavior is due to the opposing effects of impact
ionization and carrier heating, both caused by the THz pump pulse.
Electron heating leads to a decrease in mobility due to the strong
non-parabolicity in the conduction band of  InSb \cite{weng1995}
and scattering into side valleys with lower mobility
\cite{hebling2008ultrafast}, similar to the case of GaAs discussed
in the previous section.
\begin{figure}
  \includegraphics[width=8cm]{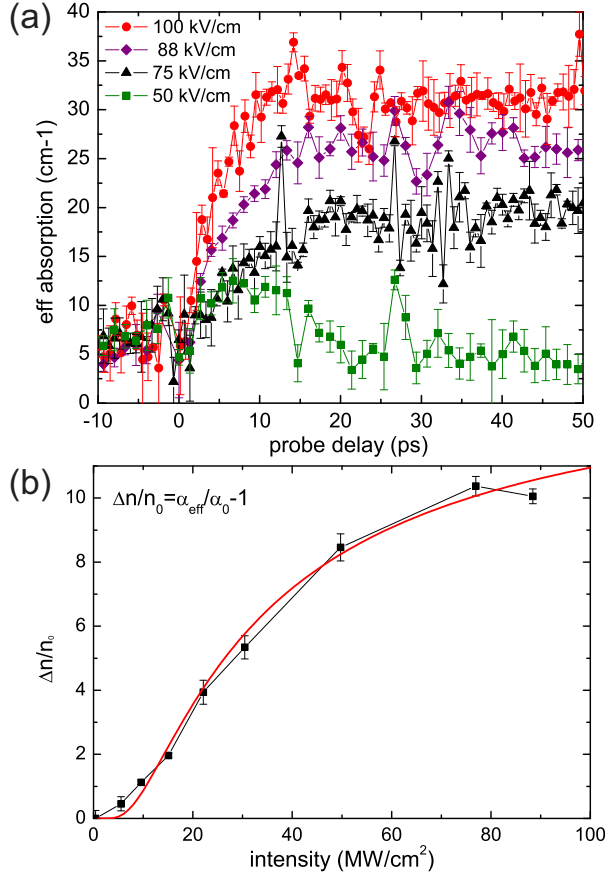}\\
  \caption{(color online) (a) time-resolved averaged absorption (0.2-1.6 THz) in undoped InSb at 80K for various fields (b) intensity dependent average absorption for the doped sample 35 ps after the THz excitation, the solid line is a fit to Eq. (2)}\label{figure6}
\end{figure}

Further pump-field dependent experimental results shown in Figure
\ref{figure6}a indicate no observable absorption increase for
single-cycle THz pulses with peak electric fields lower than 75
kV/cm in nominally undoped InSb with a carrier concentration lower
than 5$\times$10$^{14}$$^{ }$cm$^{-3}$$^{ }$at 77K.

The threshold electric field in this case should be distinguished
from that obtained in prior static or quasi-static measurements
\cite{asauskas1980}. From a simple model we were able to
extrapolate that the peak field required for impact ionization by
the single-cycle THz pulse excitation is a factor of two larger
than in the static field case \cite{hoffmann2009}. This difference
can be attributed to the significantly shorter peak pulse duration
compared to the momentum relaxation time in InSb.

The intensity dependence of  impact ionization by 40 ns
far-infrared pulse has been observed by Ganichev
\cite{ganichev1986} using photoconductive measurements.  In this
case, the electron concentration change $\Delta$n/n$_{0}$ by
impact ionization could be modeled using the Fokker-Planck
equation, leading to a dependence on the applied electric field
\textit{E} given by

\begin{equation}
\frac{\Delta n}{n}=A \exp\left(\frac{-E_0^2}{E^2}\right)
\end{equation}
where \textit{A} is a proportionality constant.

To check the validity of equation (2) in the regime of very short
pulses with high energy, we performed a series of intensity
dependent pump probe measurements with the probe delay held
constant at 35 ps after the arrival of pump. Figure 6b shows the
intensity dependent average THz absorption for the doped InSb at
80 K.  We observe a saturation of the new carrier generation at
high pump fluencies. The solid line shows fit to the experimental
data using equation (2) with parameter values \textit{A} = 14.5
and \textit{E}$_{0}$=104 kV/cm. The parameter \textit{E}$_{0}$ is
a characteristic field that gives electrons enough energy to
create an electron hole pair. While the fit gives adequate
agreement with Eq. 2, the critical field \textit{E}$_{0}$ is
roughly 2 orders of magnitude higher than that observed in Ref.
\cite{ganichev1986} for 40 ns THz laser pulses. This discrepancy
can only be understood taking into account the fundamentally
nonequilibrium carrier heating induced by the single cycle THz
pulse as opposed to the quasi-cw AC field case in previous
experiments.

\subsection{Plasma-Lattice interaction}

\begin{figure}
  \includegraphics[width=12cm]{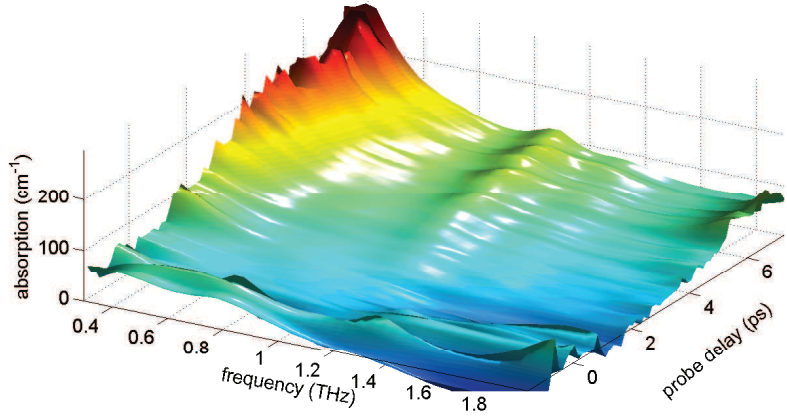}\\
  \caption{(color online) The frequency dependent absorption coefficient of InSb as a function of frequency for probe delays up to 7 ps at a temperature of 200 K.}\label{figure7}
\end{figure}

A distinct absorption peak at 1.2 THz in the undoped sample and a
weak feature that indicates a similar peak in the doped sample
were observed. Figure \ref{figure7} illustrates the appearance of
this peak for pump-probe delay times up to 7 ps in the doped
sample at 200 K. The evolution of the low frequency Drude behavior
is clearly separated from the phonon mediated peak at 1.2 THz.

The amplitude -- but not the frequency -  of the peak is highly
intensity dependent, and appears to approach its asymptotic value
just above 50\% of the maximum THz intensity. This behavior of
this peak suggests that its origin is lattice vibrational rather
than electronic. Polar optical phonon scattering is well known as
the dominant energy loss mechanism for hot electrons in InSb
\cite{conwell1967}. The main channel of energy loss is through the
emission of LO phonons with a frequency of 5.94 THz. These phonons
decay into acoustic modes through both anharmonic coupling and the
second-order electric moment of the lattice \cite{ferry1974}. A
series of sum- and difference phonon peaks between 1 to 10 THz has
been observed and assigned in the far-infrared spectrum of InSb
\cite{koteles1974}. At very low THz fields produced by a
photoconductive antenna, we were also able to observe some of
these weak absorption peaks in the undoped sample. The phonon
frequency assignments reported in \cite{koteles1974} indicate a
1.2 THz difference frequency between the LO and LA modes at the
zone boundary. The drastic change in the absorption coefficient of
the difference phonon peak is the result of large changes in
phonon populations that can be attributed to the energy transfer
from hot electrons generated by the THz pump pulse. Monte-Carlo
simulations \cite{brazis2008} have shown that substantial phonon
population changes can occur even at comparatively low DC fields
on picosecond timescales.

\section{Conclusion}

We have demonstrated the ability to accelerate free carriers in
doped semiconductors to high energies by single cycle THz pulses.
At the same time, THz probing enables the monitoring of the free
carrier absorption on the picosecond timescale. For GaAs we are
able to observe that a fraction of carriers undergo intervalley
scattering, leading to a drastic change in effective mass. In the
case of InSb, the carrier energy can exceed the impact ionization
threshold, leading to an increase of carrier concentration of
about one order of magnitude. In this material we are able to
observe the energy exchange between the hot electrons and the
lattice, leading to changes in population of the LO phonon.
Monte-Carlo simulations are needed for a more in-depth
understanding of the interplay between hot electrons and the
lattice, in which effects like impact ionization, intervalley- and
polar optical phonon scattering and changes in phonon population
are taken into account. Experimentally, a broader probe bandwidth
exceeding 2 THz will enable us to monitor the phonon dynamics
directly. With even higher THz fields we should be able to achieve
impact ionization in higher band-gap materials like Germanium or
GaAs allowing us to understand these highly nonlinear transport
phenomena without experimental complications due to avalanche
effects or multiphoton absorption.

This work was supported in part by ONR under Grant No.
N00014-06-1-0463.

\bibliography{josab_hoffmann_nelson}

\end{document}